\documentclass[3p,times,procedia]{elsarticle}
\pdfoutput=1
\usepackage{nupha_ecrc,epstopdf}

\volume{00}

\firstpage{1}

\journalname{Nuclear Physics A}

\runauth{Y.-T. Chien et al}

\jid{nupha}

\jnltitlelogo{Nuclear Physics A}

\usepackage{amssymb}

\usepackage[figuresright]{rotating}

\begin{document}

\begin{frontmatter}



\author{Yang-Ting Chien}
\address{Center for Theoretical Physics, Massachusetts Institute of Technology, Cambridge, MA 02139, U.S.A.}
\dochead{XXVIIth International Conference on Ultrarelativistic Nucleus-Nucleus Collisions\\ (Quark Matter 2018)}

\title{Probing heavy ion collisions using quark and gluon jet substructure with machine learning}


\begin{abstract}
Understanding the inner working of the quark-gluon plasma requires complete and precise jet substructure studies in heavy ion collisions.
In this proceeding we discuss the use of quark and gluon jets as independent probes, and how their classification allows us to uncover regions of QCD phase space sensitive to medium dynamics. We introduce the telescoping deconstruction (TD) framework to capture complete jet information and show that TD observables reveal fundamental properties of quark and gluon jets and their modifications in the medium. We draw connections to soft-drop subjet distributions and illuminate medium-induced jet modifications using Lund diagrams. The classification is also studied using a physics-motivated, multivariate analysis of jet substructure observables. Moreover, we apply image-recognition techniques by training a deep convolutional neural network on jet images to benchmark classification performances. We find that the quark gluon discrimination performance worsens in \textsc{Jewel}-simulated heavy ion collisions due to significant soft radiation affecting soft jet substructures. This work suggests a systematic framework for jet studies and facilitates direct comparisons between theoretical calculations and measurements in heavy ion collisions.
\end{abstract}


\end{frontmatter}


\section{Introduction}
\label{intro}

\begin{figure}
    \includegraphics[height=2.4cm, trim = {0mm 0mm 0mm 0cm}, clip]{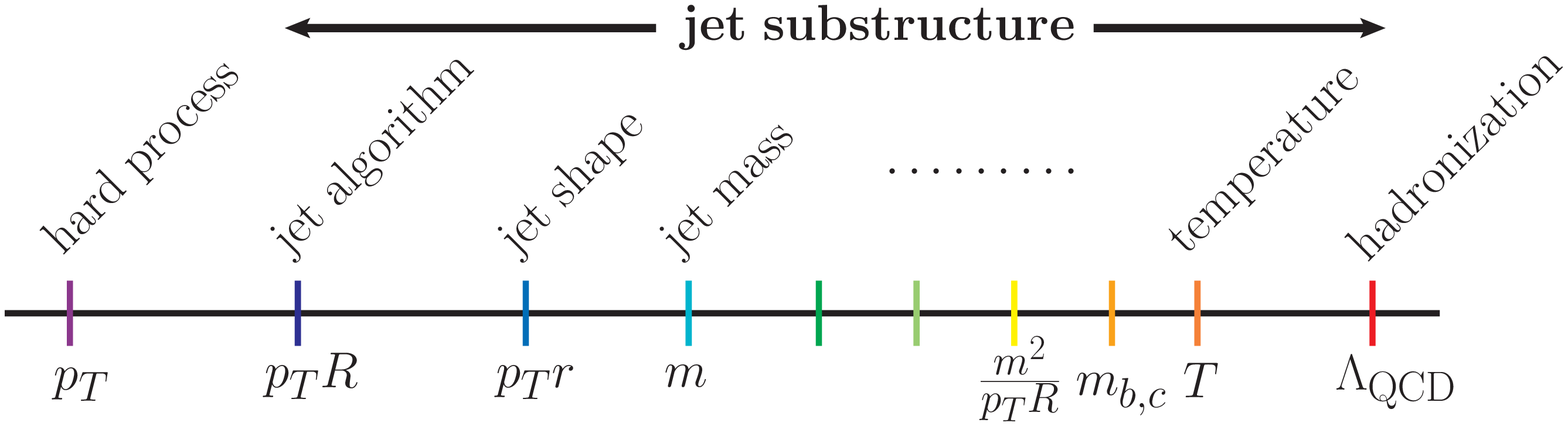}
    \includegraphics[height=2.6cm, trim = 0mm 0cm 0mm 0cm , clip=true]{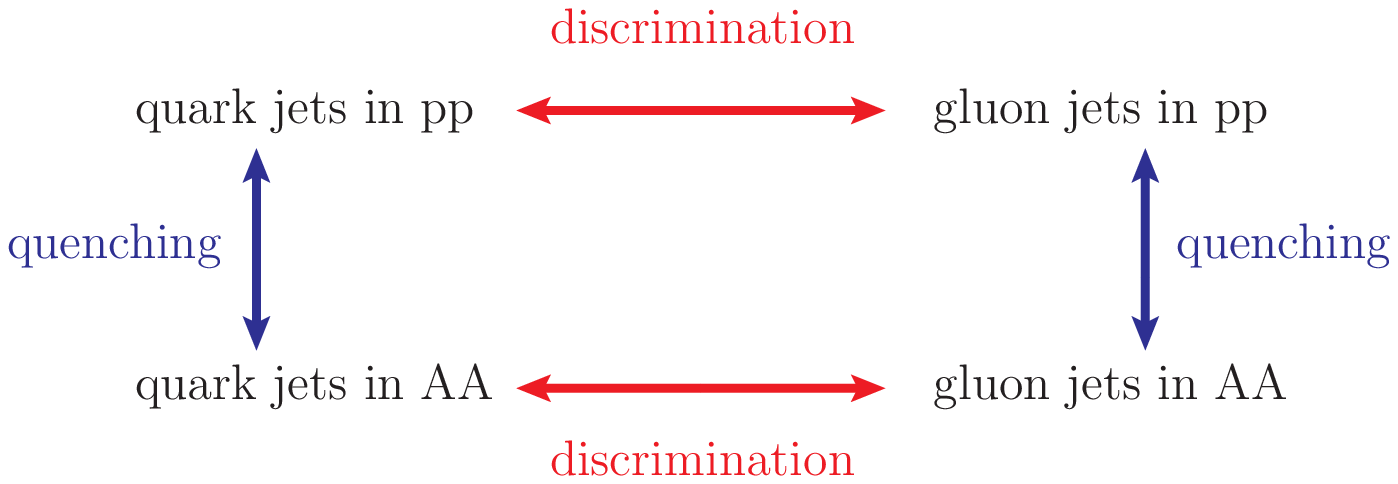}
\caption{Left panel: Jet substructure observables can probe QCD dynamics in all energy regimes from the highest scale down to $\Lambda_{\rm QCD}$ \cite{Chien:2015hda}. Right panel: Classification of quark and gluon jets in pp and AA collisions provide a new method for jet quenching studies \cite{Chien:2018dfn}.}
    \label{fig:sub_scale}
\end{figure}

The study of jet quenching has moved onto detailed analysis of the redistribution of jet energy quantified by jet substructure modifications. It has been realized that different jet substructure observables are sensitive to different underlying QCD dynamics at characteristic energy scales (left panel of Fig. \ref{fig:sub_scale}). One can design jet substructure observables to probe specific regions of phase space where jet-medium interaction may have the dominant effect. A comprehensive examination of jet substructure modifications will then allow us to search for possible signatures which may reveal fundamental properties of the quark-gluon plasma (QGP). On the other hand, a change of quark and gluon jet fractions in heavy ion collision can contribute significantly to jet substructure modifications. An increase of the quark-jet fraction due to larger suppressions of gluon jets can make the jet energy profile more quark-jet like, an important effect in addition to the jet-by-jet modification to substructure \cite{Chien:2015hda,Chatrchyan:2013kwa}. This further motivates the studies of jet modifications with different quark and gluon jet fractions which enable the use of quark and gluon jets as independent probes.

\begin{figure}
    \includegraphics[height=4.23cm, trim = {0mm 0mm 0mm 0cm}, clip]{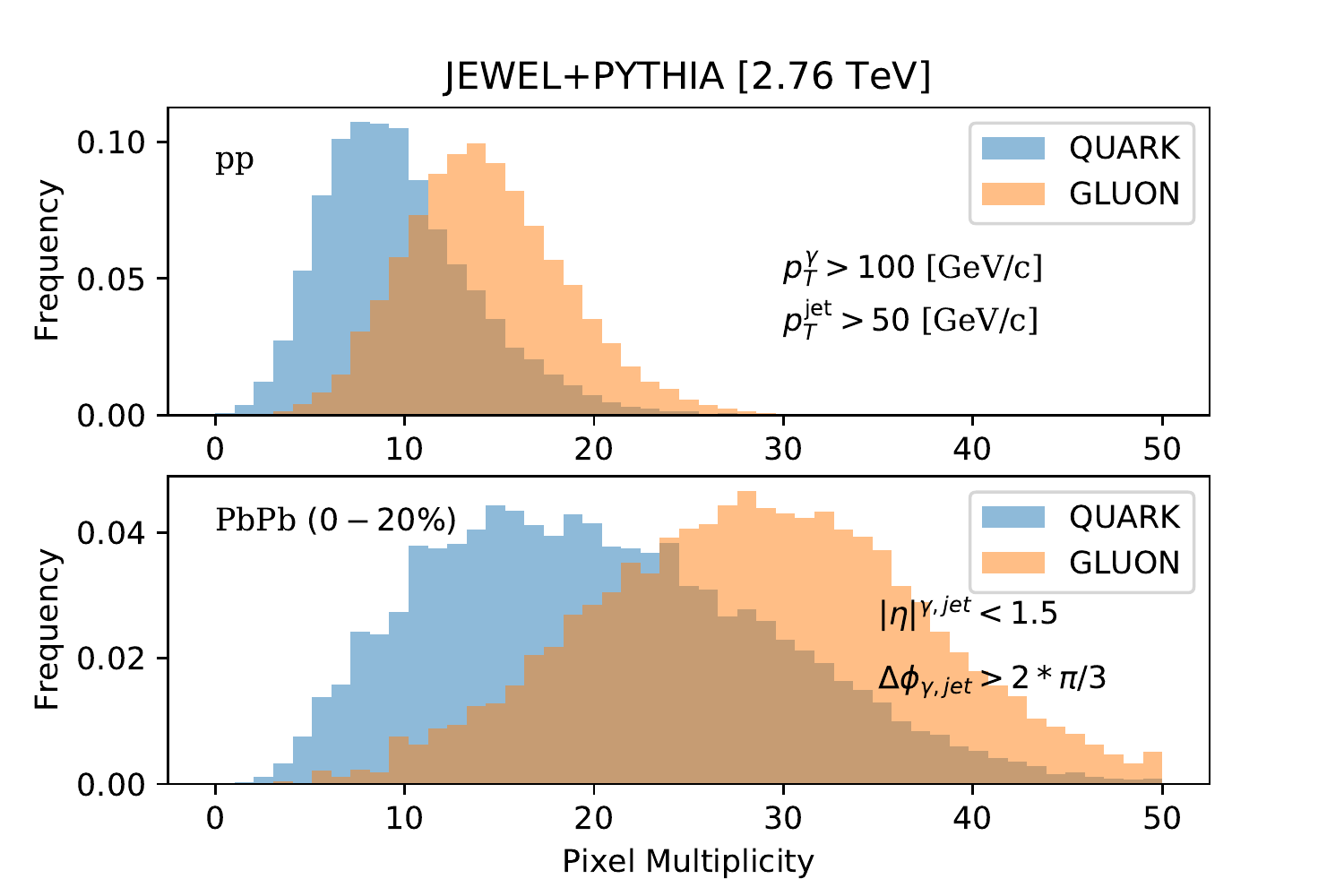}
    \includegraphics[height=3.92cm, trim = {0mm 0mm 0mm 0cm}, clip]{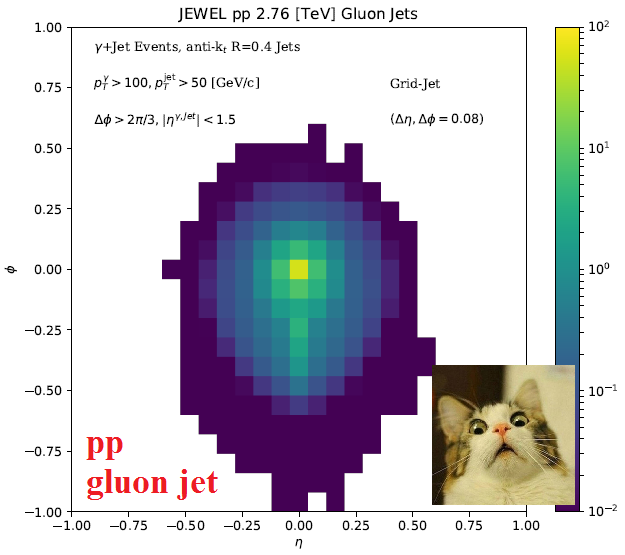}
    \includegraphics[height=3.92cm, trim = {0mm 0mm 0mm 0cm}, clip]{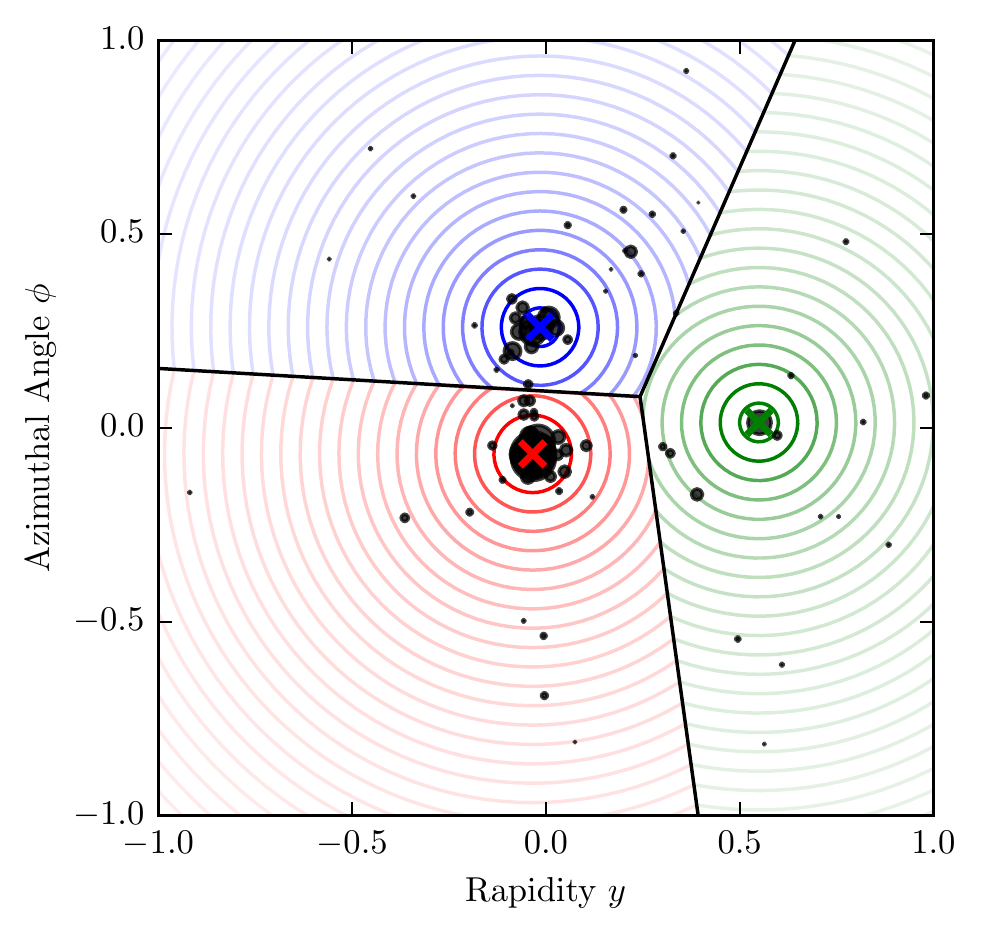}
\caption{Left panel: Pixel multiplicity distributions for quark and gluon jets in $pp$ (upper) and $AA$ (lower) collisions simulated using \textsc{Jewel}. Middle panel: Average gluon jet image in $pp$ collisions. Right panel: Telescoping deconstruction of a QCD jet at the T3 order.}
\label{fig:three}
\end{figure}

In this proceeding we exploit this idea and study classifications of quark and gluon jets in $pp$ and $AA$ collisions. The goal is to extract complete jet features which encode all aspects of jet modifications in $AA$ collisions. We study the discrimination of jets in $pp$ and $AA$ collisions and show that it is intimately related to quark gluon discrimination (right panel of Fig. \ref{fig:sub_scale}) which aims to identify differences between quark and gluon jets. We use three approaches, starting from a multivariate analysis of a list of physics-motivated jet observables (left panel of Fig. \ref{fig:three}). On the other hand, we apply image recognition techniques which identify relevant features using machine learning methods (middle panel of Fig. \ref{fig:three}). In between we introduce the telescoping deconstruction framework which aims to organize and capture complete physical information within jets using telescoping subjets (right panel of Fig. \ref{fig:three}). Below we briefly summarize each of the method.

\section{Quark and gluon jet substructure and modification}
\label{qgsub}

The quark and gluon enriched jet samples used in this work were generated using the prompt photon production channels $q+\gamma$ and $g+\gamma$ in \textsc{Jewel}. The physics-motivated, multivariate analysis constructively combines information captured in each individual jet observable. We consider five representative ones: jet mass and radial moments which are infrared and collinear (IRC) safe, as well as IRC-unsafe observables of $p_T^D$ and pixel multiplicity. In general, from individual plots we see that gluon jets have broader energy distributions and softer hadron fragmentation compared to quark jets, and medium interactions result in broader energy distribution and softer hadron fragmentation for both quark and gluon jets. The jet image method trains a deep convolutional neural network (CNN) on quark and gluon jet images in $pp$ and $AA$ collisions \cite{Komiske:2016rsd}. The energy distribution in rapidity $y$ and azimuthal angle $\phi$ is discretized with a finite pixel size. The CNN is then a powerful model capable of processing raw pixel jet data and find useful features which help maximize the separation among jet samples. From the average jet images, we see again that gluon jets are more spread out and populating more pixels with soft particles compared to quark jets, and the medium broadens the energy distribution.

\begin{figure}
    \includegraphics[width=.245\columnwidth, trim = {0mm 0mm 0mm 0cm}, clip]{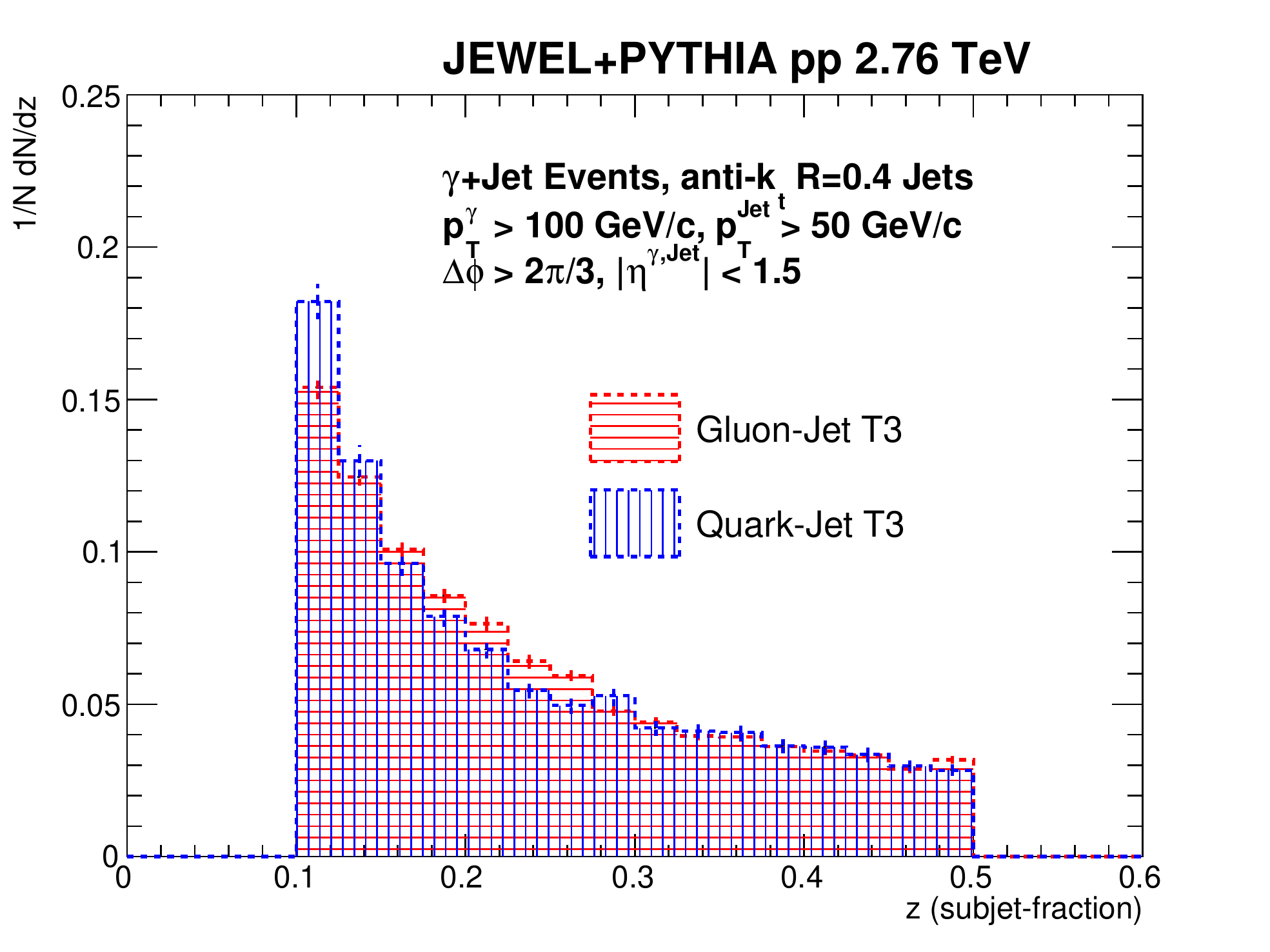}
    \includegraphics[width=.245\columnwidth, trim = {0mm 0mm 0mm 0cm}, clip]{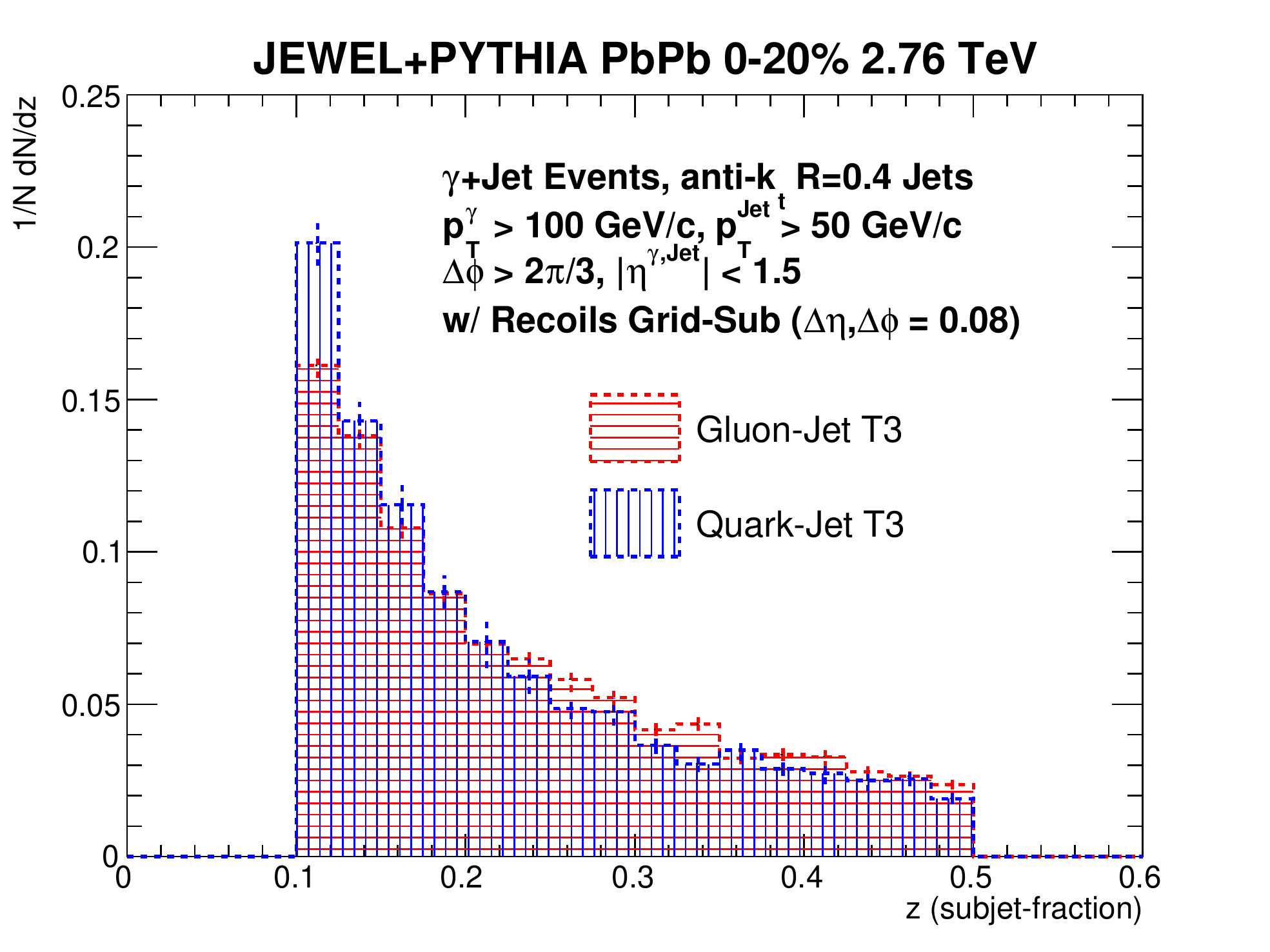}
    \includegraphics[width=.245\columnwidth, trim = 0mm 0cm 0mm 0cm , clip=true]{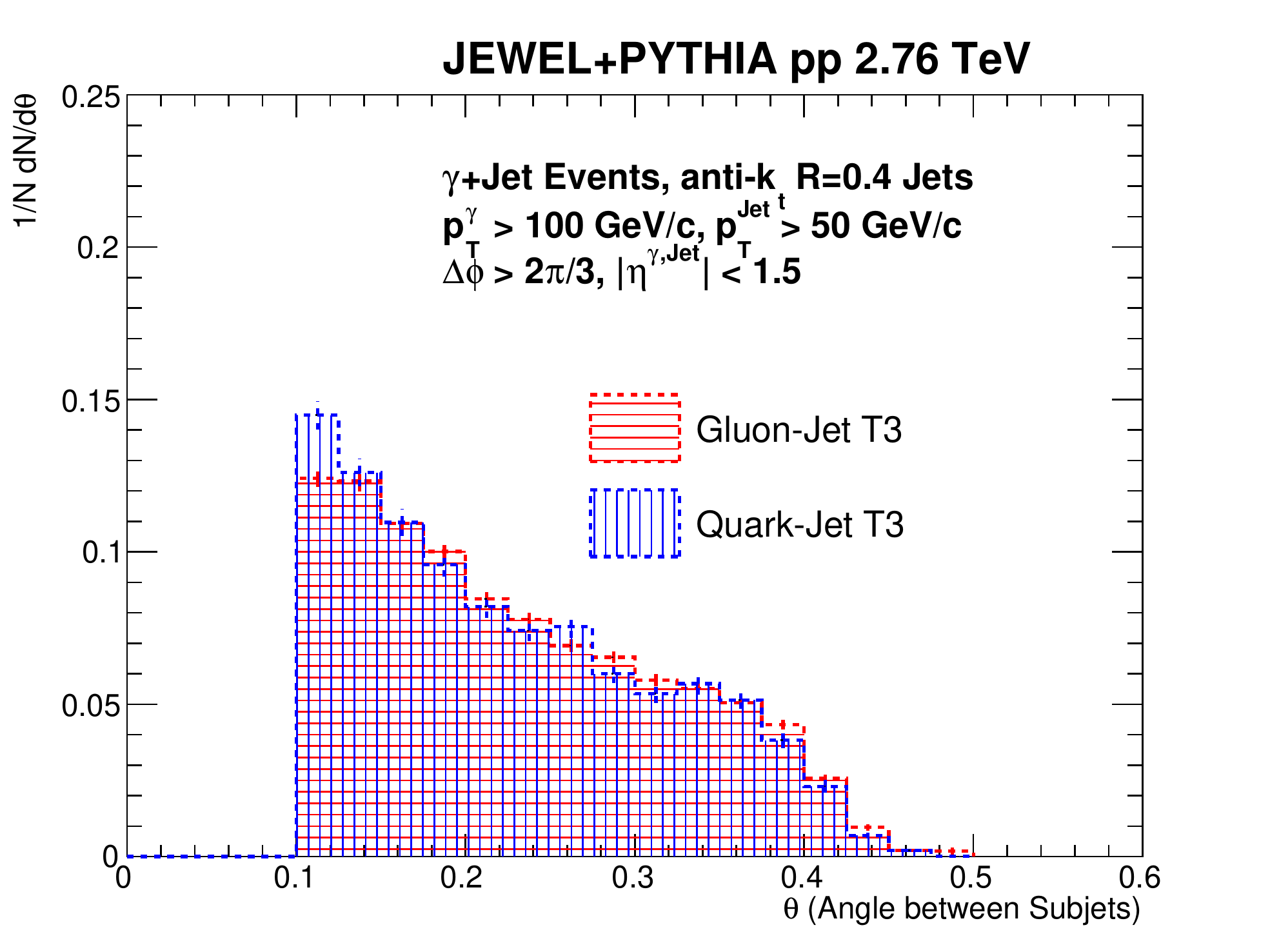}
    \includegraphics[width=.245\columnwidth, trim = 0mm 0cm 0mm 0cm , clip=true]{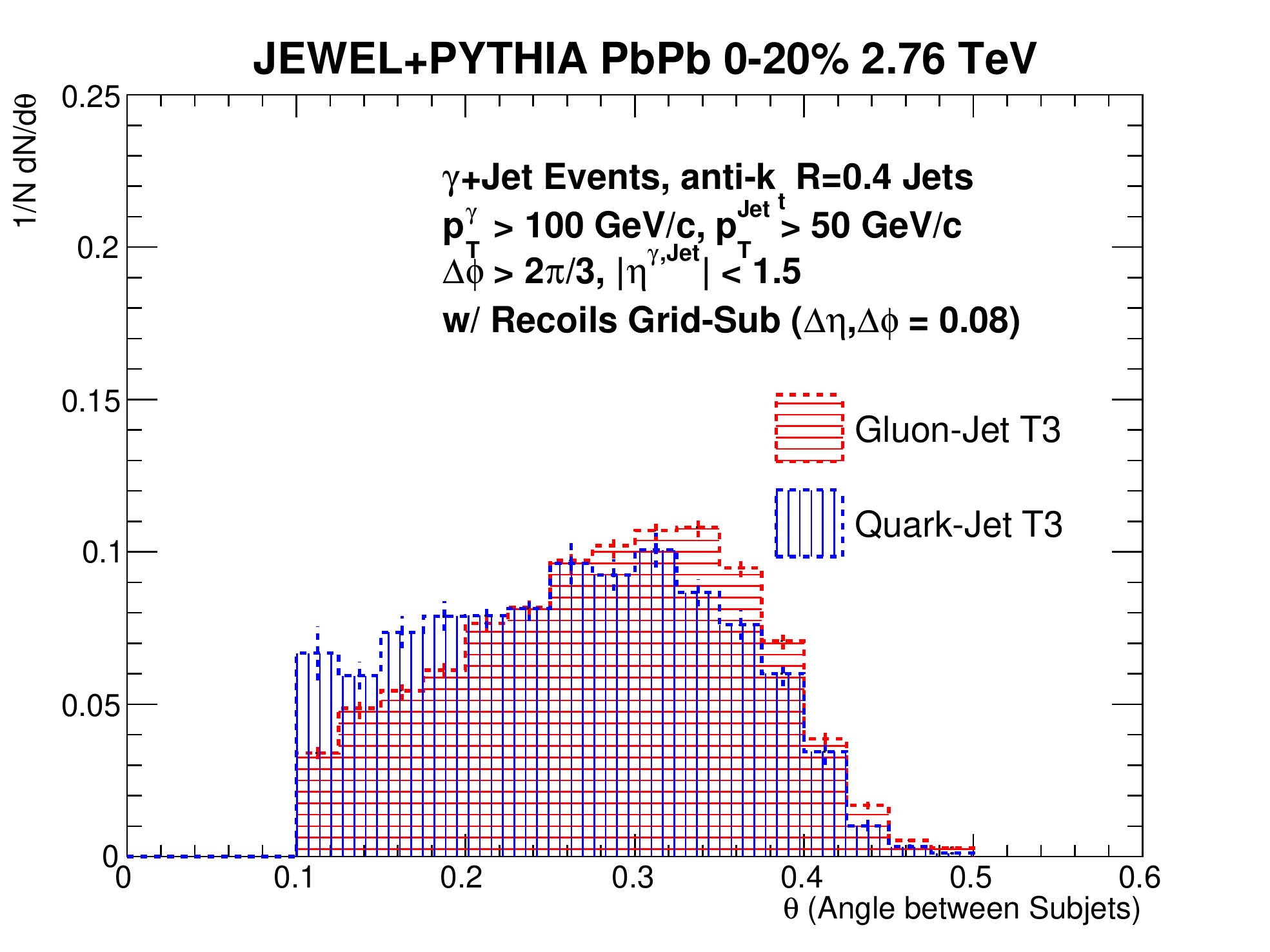}
\caption{Left two panels: Subjet momentum fraction distributions at the T3 order for quark and gluon jets in $pp$ (left) and $AA$ (right) collisions. Right two panels: Subjet angular distributions at the T3 order for quark and gluon jets in $pp$ (left) and $AA$ (right) collisions.}
\label{fig:z_theta}
\end{figure}

\begin{figure}
    \includegraphics[width=.234\columnwidth, trim = {0mm 0mm 0mm 0cm}, clip]{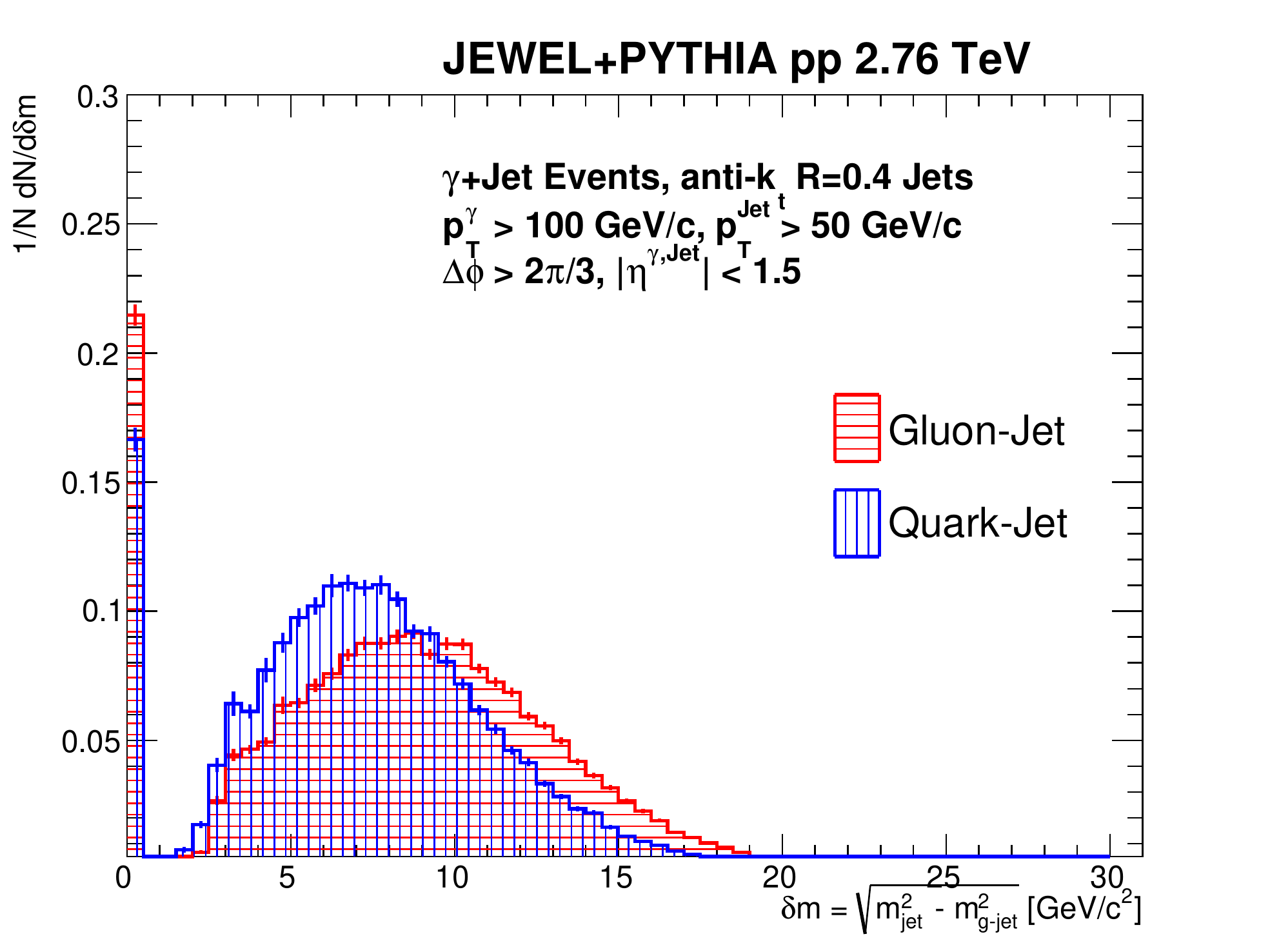}
    \includegraphics[width=.234\columnwidth, trim = {0mm 0mm 0mm 0cm}, clip]{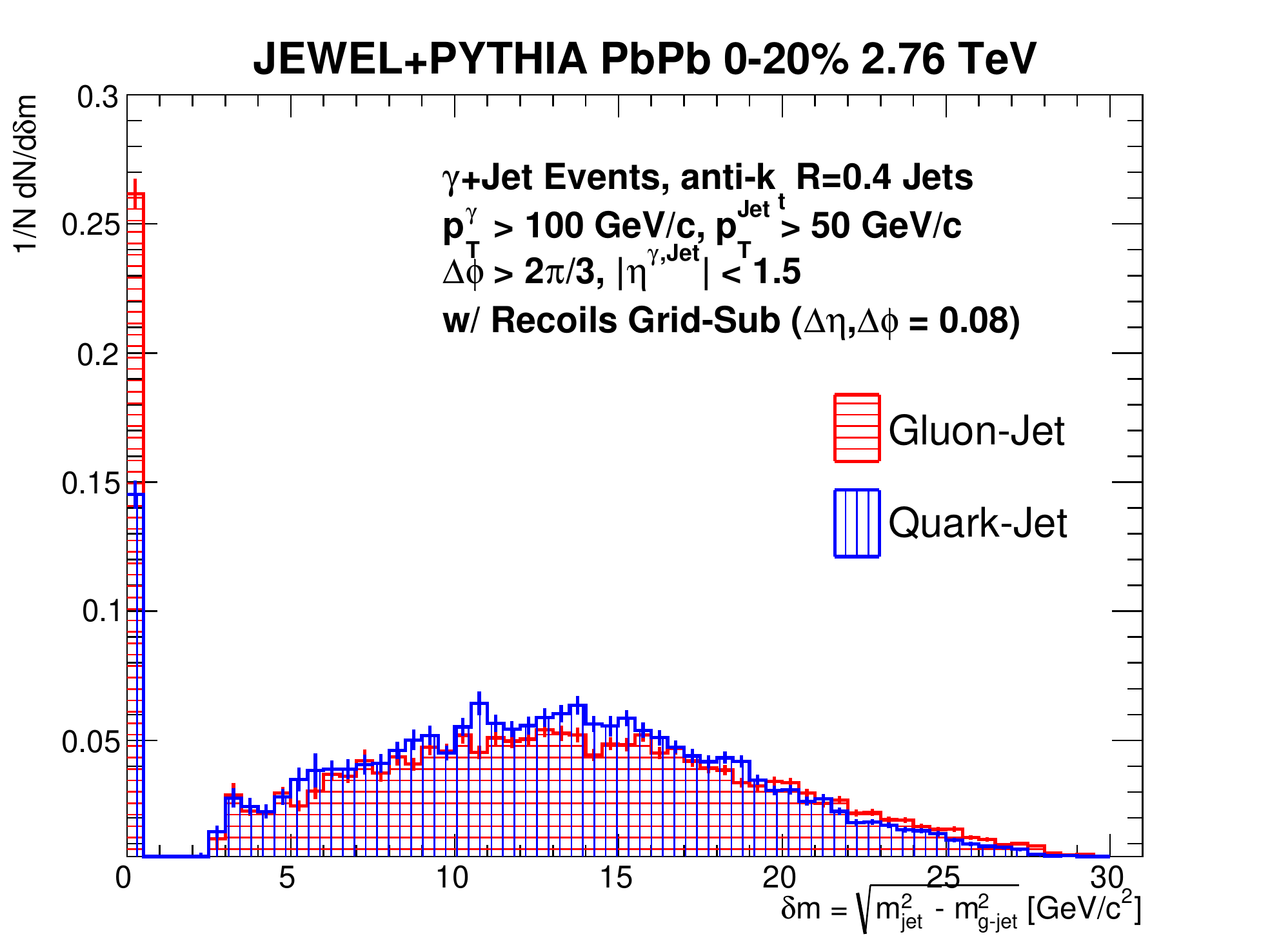}
    \includegraphics[width=.258\columnwidth, trim = {0mm 0mm 0mm 0cm}, clip]{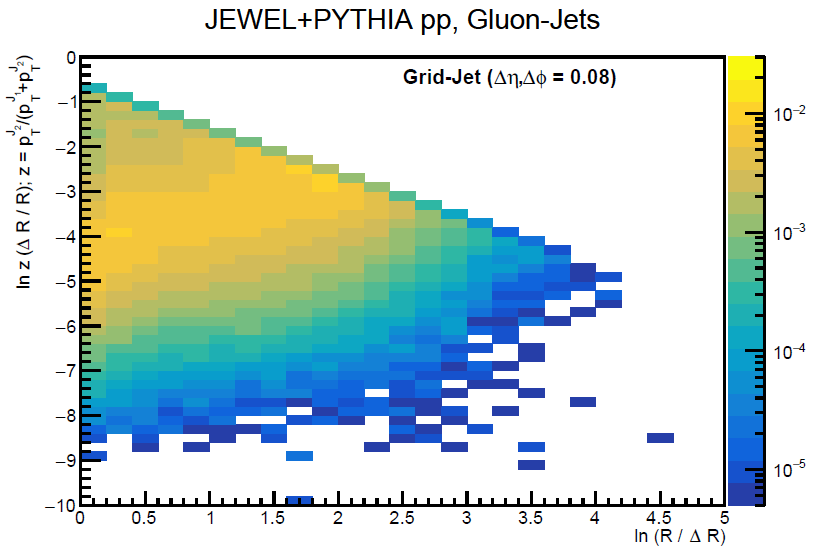}
    \includegraphics[width=.258\columnwidth, trim = {0mm 0mm 0mm 0cm}, clip]{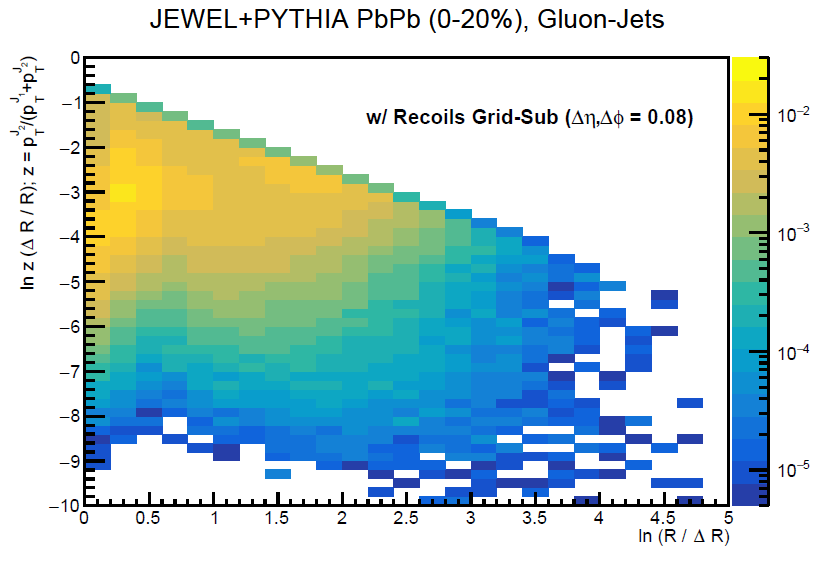}
\caption{Left two panels: Distributions of the collinear-drop observable $\delta m$ for quark and gluon jets in $pp$ (left) and $AA$ (right) collisions. Right two panels: Lund planes for gluon jets in $pp$ (left) and $AA$ (right) collisions.}
\label{fig:cd_lund}
\end{figure}

The TD framework probes energy flows within jets using subjets with multiple angular resolutions \cite{Chien:2013kca,Chien:2017xrb}. It decomposes jet information in a fixed-order expansion organized by the number of reconstructed subjets. The procedure starts from identifying $N$ dominant energy flow directions along soft-recoil free axes. Exclusive subjets are then reconstructed around the axes with multiple subjet radii $R_T$, and subjet kinematic variables form a complete jet substructure basis. We show that subjet momentum fraction $z$ and angular distributions $\theta$ constructed in telescoping deconstruction encode fundamental QCD properties such as the Altarelli-Parisi splitting functions (Fig. \ref{fig:z_theta}), similar to the groomed momentum sharing $z_g$ and groomed jet radius $r_g$ constructed in Soft Drop \cite{Larkoski:2014wba}. Note the characteristic $1/z$ functional form in the subjet momentum fraction. Recently the soft-drop $z_g$ variable was used to probe heavy ion collisions with significant enhancement of soft subjets \cite{Sirunyan:2017bsd} which was first explained as a signature of medium-induced radiation \cite{Chien:2016led}. We see a similar modification pattern in the $z$ distribution and also show that the $\theta$ distribution receives strong medium modifications enhancing wide-angle emissions. To go beyond we examine subjet masses which further reveal the flavor origin of quark and gluon jets, with significant modification in $AA$ collisions hinting at a large, soft radiation component. This is further tested using a collinear-drop observable $\delta m$ which is designed to probe soft radiation within jets (left two panels of Fig. \ref{fig:cd_lund}). We see that the difference between quark and gluon jets in $\delta m$ disappears in $AA$ collisions which suggests that soft radiation washes out such feature which distinguish quark and gluon jets, a possible signature of medium response to jets.

The two-dimensional, longitudinal and transverse distribution of subjets can be represented and studied using the Lund plane \cite{Andersson1989} (right two panels of Fig. \ref{fig:cd_lund}), where sequential branching kinematics of each jet is recorded. The infrared properties of QCD implies a uniform distribution in the Lund plane, with effect from running coupling and hadronization affecting the soft region. Comparing Lund planes in $pp$ and $AA$ collisions, we see a clear enhancement of soft, wide angle branches which is consistent with the modification patterns seen in previous jet substructure observables. We also look into how Soft Drop affects the Lund plane distributions and observe that both hard and soft branches are modified, which is consistent with the modifications of subjet masses seen in telescoping deconstruction.

\section{Quark and gluon jet classification}
\label{qgdis}

\begin{figure}
    \centering
    \includegraphics[width=.42\columnwidth, trim = {0mm 0mm 0mm 0cm}, clip]{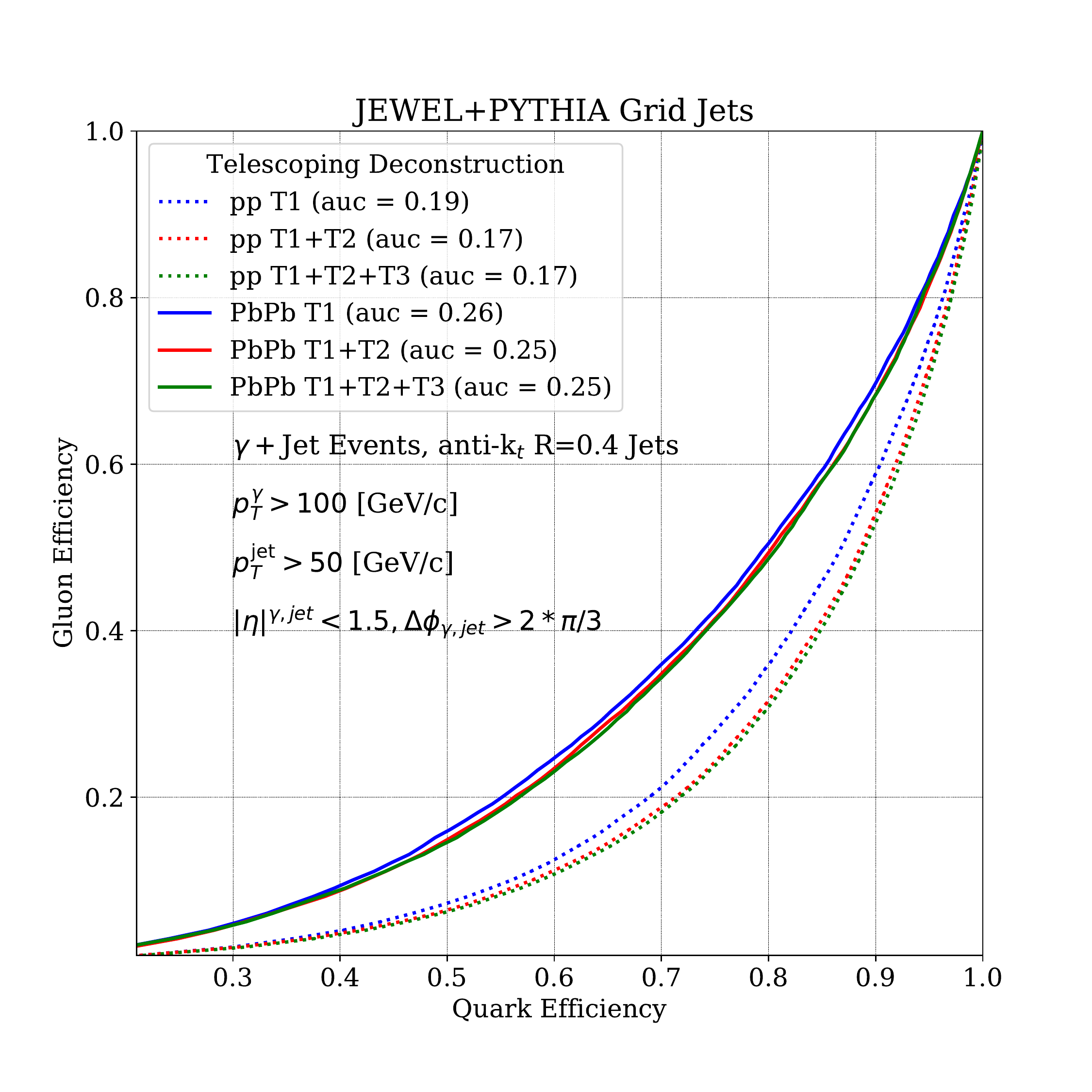}~~~~~
    \includegraphics[width=.42\columnwidth, trim = {0mm 0mm 0mm 0cm}, clip]{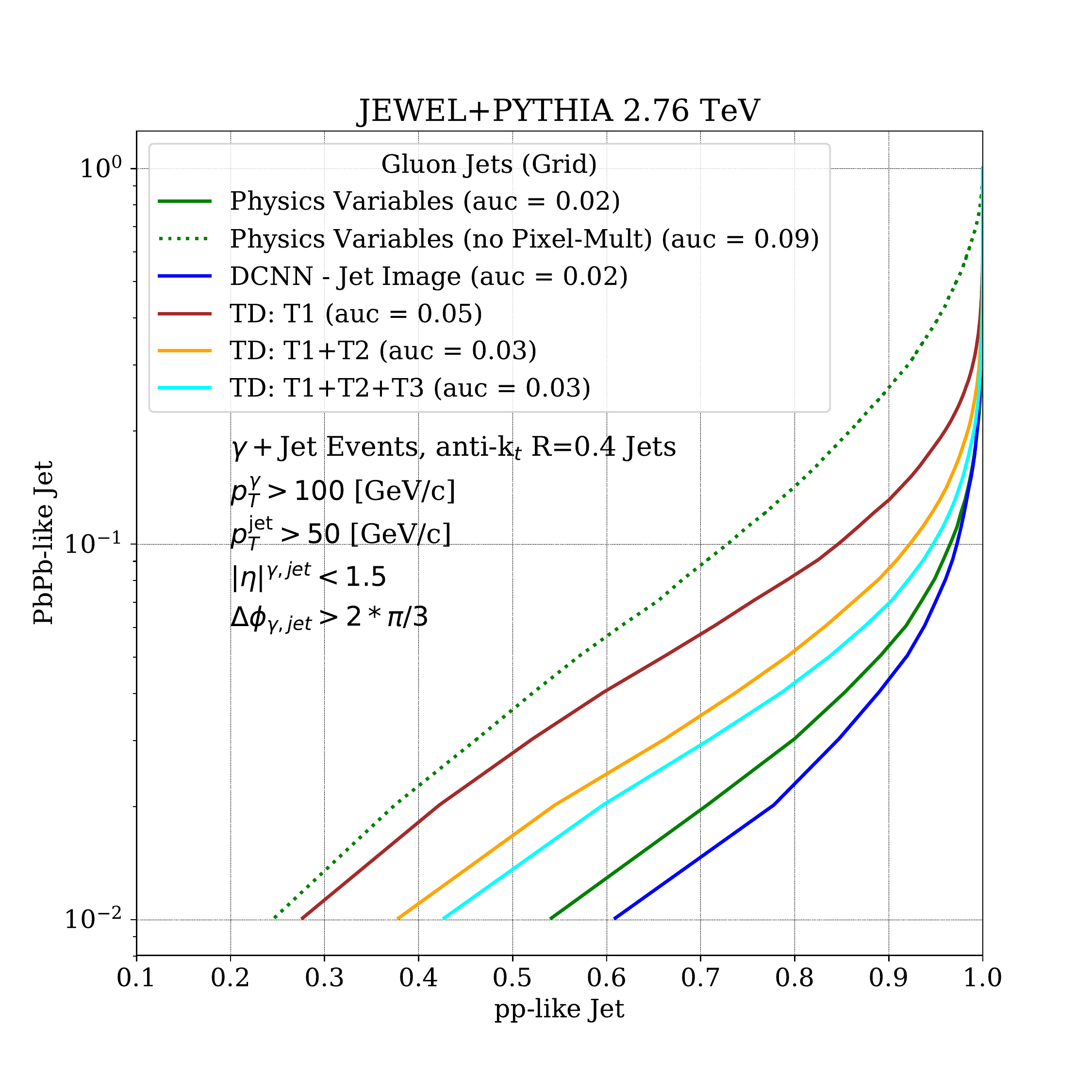}
\caption{Left panel: ROC curves using TD variables for quark gluon discrimination in $pp$ and $AA$ collisions. Right panel: ROC curves using physics-motivated multivariate analysis, jet image and TD for discriminating gluon jets in $pp$ and $AA$ collisions. }
\label{fig:roc}
\end{figure}

Having examined the jet substructure information represented using physics-motivated observables, jet images and TD basis, we combine all the information in each category using multivariate analysis tools and study the classification of quark and gluon jets in $pp$ and $AA$ collisions. A proper neural network architecture is chosen for processing the simulated input data. We perform two tasks, discriminating quark jets and gluon jets, and discriminating jets in $pp$ and $AA$ collisions. We quantify the classification performance using receiver-operating-characteristic (ROC) curves, plotting signal efficiency versus background efficiency (Fig. \ref{fig:roc}) with higher performance towards the lower-right corner of the plots. We see that all methods give consistent and comparable performance, suggesting that with the discretization resolution each method captures most of the substructure information. The telescoping deconstruction performance converges quickly with increasing T$N$ order. We find that the quark gluon discrimination performance goes down in \textsc{Jewel}-simulated $AA$ collisions. Also, the pixel multiplicity is the dominant observable distinguishing jets in $pp$ and $AA$ collisions, a characteristic feature of the significant soft event activities.

\section{Conclusions}
\label{conc}

Jet substructure observables encode information about jet-medium interaction. We show that quark and gluon jet substructure can be independent probes and that quark gluon discrimination is a new way for jet modification studies. We use physics-motivated multivariate analysis and machine learning tools, and we develop the TD framework to decompose jet information using subjet basis. We emphasize the importance of comprehensive substructure studies which may lead to the understanding of the inner working of QGP.





\bibliographystyle{elsarticle-num}
\bibliography{QMref}







\end{document}